\documentclass{aip-cp}

\usepackage[numbers]{natbib}
\usepackage{rotating}
\usepackage{graphicx}
\usepackage[abs]{overpic}

\begin{document}

\title{Exclusive Diffractive Resonance Production in Proton-Proton Collisions at the LHC}

\author[aff1]{Rainer Schicker\corref{cor1}}
\author[aff2]{Roberto Fiore}
\eaddress{fiore@cs.infn.it}
\author[aff3]{Laszlo Jenkovszky}
\eaddress{jenk@bitp.kiev.ua}

\affil[aff1]{Physikalisches Institut, Im Neuenheimer Feld 226, 
Heidelberg University, 69120 Heidelberg, GERMANY}
\affil[aff2]{Department of Physics, University of Calabria,
I-87036 Arcavacata di Rende, Cosenza, ITALY}
\affil[aff3]{Bogolyubov Institute for Theoretical Physics (BITP),
Ukrainian National Academy of Sciences 14-b, Metrologicheskaya
str., Kiev, 03680, UKRAINE}
\corresp[cor1]{Corresponding author: schicker@physi.uni-heidelberg.de}

\maketitle

\begin{abstract}
A model for exclusive diffractive resonance production in proton-proton 
collisions at LHC energies is presented. This model is based on the 
convolution of the Donnachie-Landshoff parameterisation of Pomeron flux in the 
proton with the Pomeron cross section for resonance production. The 
hadronic cross section for f$_{0}$(980) and f$_{2}$(1270) production at 
midrapidity is given differentially in mass and transverse momentum  of the 
resonance. The proton fractional longitudinal momentum loss is presented.    

\end{abstract}

\section{INTRODUCTION}

Central production has been studied at the energies $\sqrt{s}$ = 12.7-63 GeV 
of the Intersecting Storage Ring (ISR) at CERN, at the SPS by the COMPASS 
Collaboration, at the Tevatron by the CDF Collaboration, at RHIC by
the STAR Collaboration, and at the LHC by the ALICE and LHCb Collaborations.
At all these energies, pronounced resonance structures are seen in the two-pion 
invariant mass spectra at values \mbox{m$_{\pi\pi} < $ 2 GeV/$c^{2}$.} 
The analysis of central production events recorded by complex detector systems 
necessitates the simulation of such events by Monte Carlo generators. The 
purpose of the study presented here is the formulation of such an event 
generator in the low mass resonance region m$_{\pi\pi} <$ 2 GeV/$c^{2}$ where the 
perturbative QCD formalism is not applicable.

\section{CENTRAL PRODUCTION AT HADRON COLLIDERS}

\begin{figure*}
\vspace{-0.7cm}
\includegraphics[width=.31\textwidth]{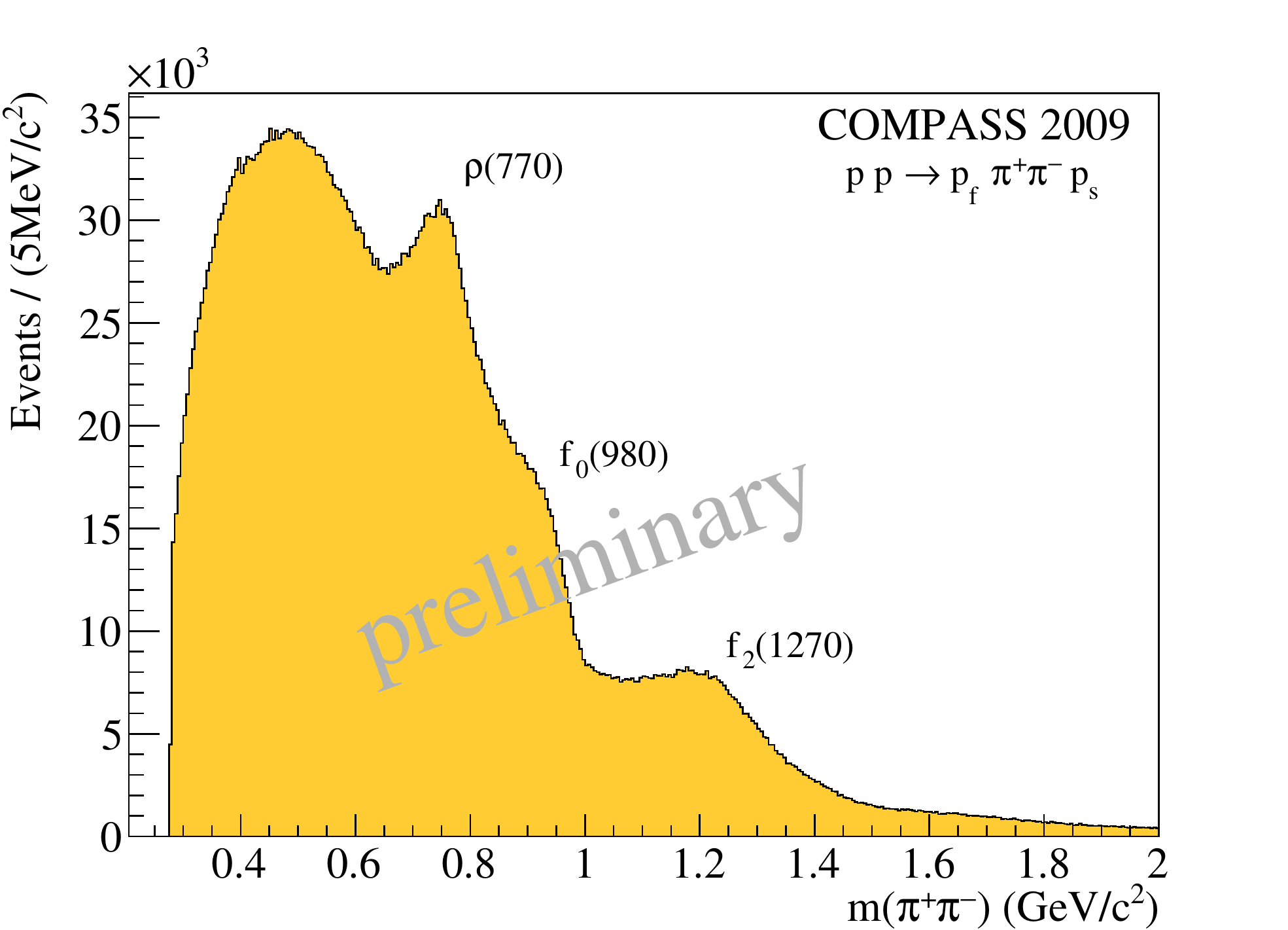}
\hspace{-0.3cm}
\includegraphics[width=.352\textwidth]{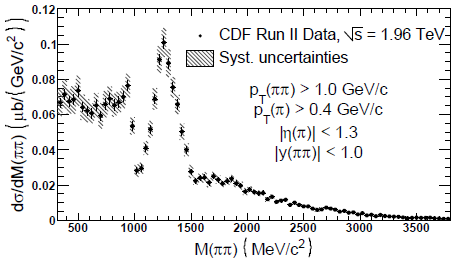}
\hspace{0.3cm}
\includegraphics[width=.305\textwidth]{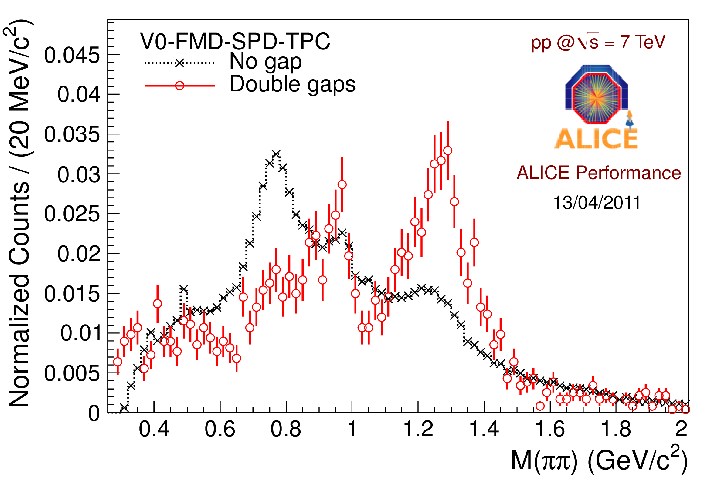}
\caption{Invariant pion pair masses from the COMPASS Collaboration on the left, 
the CDF Collaboration in the middle, and the ALICE Collaboration on the right.}
\label{fig1}
\end{figure*}

The pion-pair invariant mass spectra measured by the 
COMPASS \cite{COMPASS_pair}, the CDF \cite{CDF_pair} and the 
\mbox{ALICE Collaboration \cite{ALICE_pair}} are shown in Figure \ref{fig1}. 
Clearly visible in all the three spectra
are the f$_{2}$(1270) resonance, and the $\rho$(770) and f$_{0}$(980)

\hspace{-.7cm}resonance in the COMPASS data. The invariant mass 
spectra of the CDF and the ALICE measurements are seriously distorted at 
low masses due to the single track p$_{T}$-threshold. These mass spectra 
contain only the pion pairs with transverse momenta of approximately p$_{T}>$ 
0.5 and 0.8 GeV/$c$ for the ALICE and CDF measurements, respectively.
The missing acceptance of resonances in the low-p$_{T}$ region 
can, however, be modelled with the help of an event generator tuned
to the visible part of the p$_{T}$-spectrum.   

\section{EXCLUSIVE DIFFRACTIVE RESONANCE PRODUCTION}

The model for exclusive resonance production is developed in two steps. In the 
first step, the amplitude for Pomeron-Pomeron scattering to mesonic states is 
formulated, and the corresponding cross section is derived \cite{PomPom}.
In the second step, this Pomeron-Pomeron cross section is convoluted
with the Pomeron distribution in the proton in order to get
the resonance production cross section at hadron level.

\subsection{The Dual Resonance Model of Pomeron-Pomeron Scattering}

The amplitude for Pomeron-Pomeron scattering to mesonic states is defined by 
the dual amplitude with Mandelstam analyticity (DAMA) \cite{DAMA}. From the 
direct-channel pole decomposition of this amplitude, the cross section of 
Pomeron-Pomeron scattering producing low-mass resonances is derived by use of 
the optical theorem. The Pomeron-Pomeron channel, 
$PP \rightarrow M_{X}^{2}$, has couplings to the Pomeron and $f$ channels as 
defined by conservation of \mbox{quantum numbers \cite{PomPom}.}  

\begin{figure*}[h]
\begin{center}
\vspace{-0.4cm}
\includegraphics[width=.56\textwidth]{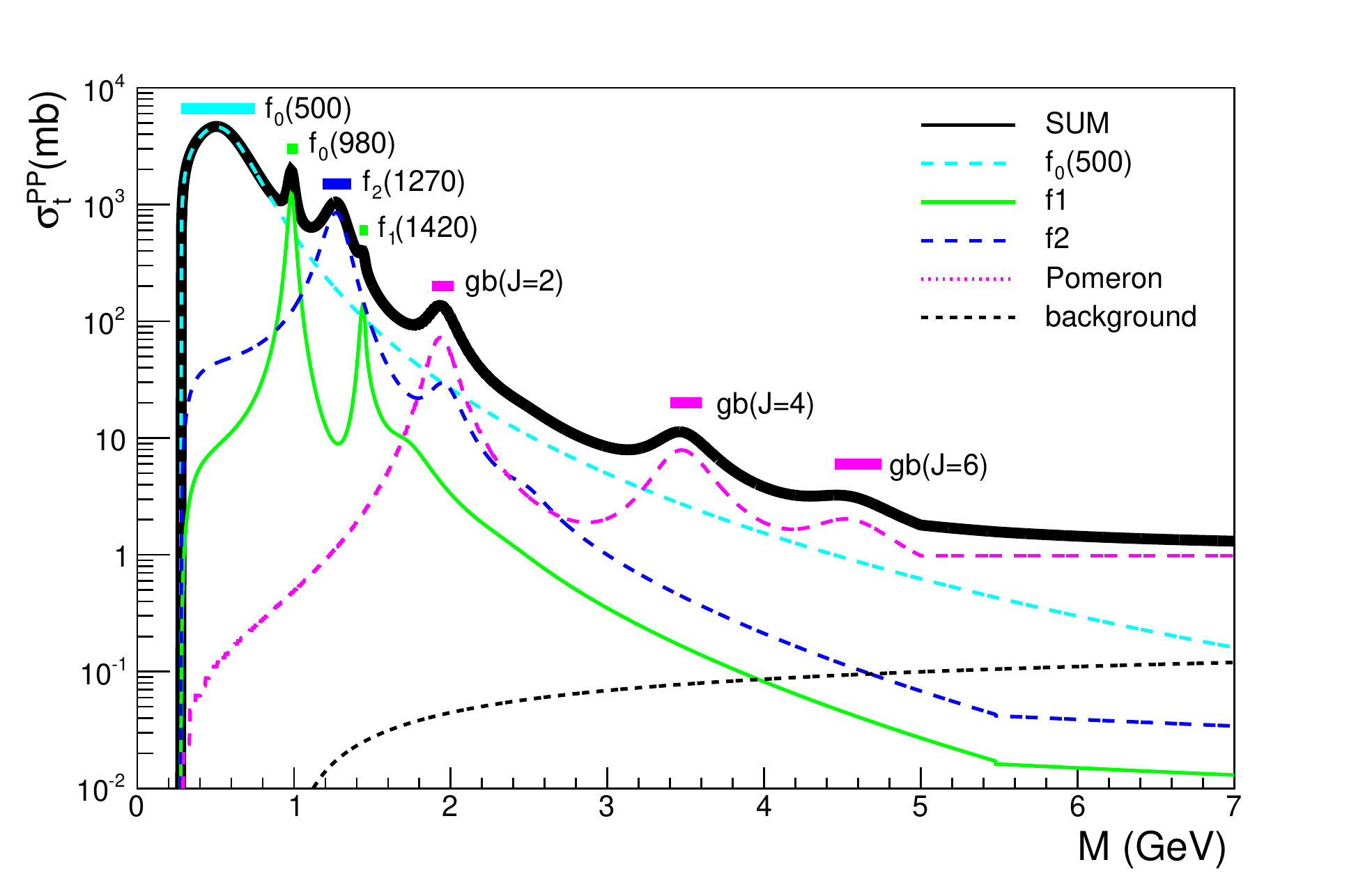}
\caption{The Pomeron-Pomeron total cross section as function of the 
resonance mass (figure from Ref. \cite{PomPom}).}
\label{fig2}
\end{center}
\end{figure*}

In Figure \ref{fig2}, the contributions  of the f$_{0}$(500) resonance, the 
$f_{1}$, $f_{2}$ and the Pomeron trajectory, and of the background to the 
Pomeron-Pomeron total cross section are shown. The contribution of the 
f$_{0}$(500) resonance, indicated by the dashed cyan line, is calculated by 
taking the central values for the mass and the width of this resonance,
$M_{0} = 475\,$MeV and $\Gamma = 550\,$MeV, respectively. The contribution of 
the $f_{1}$ trajectory shown by the solid green line clearly shows the 
f$_0$(980) resonance. The contribution of the $f_{2}$ trajectory, displayed 
by the dashed  blue line, shows peaks for the f$_2$(1270) and the 
f$_4$(2050) resonances.

\subsection{The Cross Section at Hadron Level}

The cross section at hadron level can be derived by using the definition of 
a cross section element

\begin{equation}
d\sigma = \frac{|\mathcal{M}|^{2}}{\textrm{flux}} dQ,  
\label{eq1}
\end{equation}

\noindent with $\mathcal{M}$ the invariant amplitude of the process, $dQ$ the 
Lorentz-invariant phase space, and the flux in the denominator representing 
the flux factor. Equation (\ref{eq1}) can be rewritten as 

\begin{equation}
|\mathcal{M}|^{2} dQ = \textrm{flux}_{prot}\; d\sigma_{prot} = 
\textrm{flux}_{Pom} \; \textrm{F}^{Pom}_{prot} \; d\sigma_{Pom}.
\label{eq2}
\end{equation}
   
In Equation (\ref{eq2}), the quantity $\textrm{F}^{Pom}_{prot}$ represents
the distribution of Pomerons in the proton. The flux factor for collinear 
two-body collision of particle A and B is given by \cite{QCDBook}

\begin{equation}
\textrm{flux} = 4.*\big( (p_{A}\cdot p_{B})^{2}-m_{A}^{2} m_{B}^{2}\big)^{1/2}.
\label{eq3}
\end{equation}

\noindent From Equation (\ref{eq2}), a cross section element at hadron level 
is defined as 

\begin{equation}
d\sigma_{prot} = \frac{\textrm{flux}_{Pom}}{\textrm{flux}_{prot}} 
\; \textrm{F}^{Pom}_{prot} \; d\sigma_{Pom}.
\label{eq4}
\end{equation}
  
\noindent The distribution of Pomerons in the proton, $\textrm{F}^{Pom}_{prot}$, 
can be expressed as function of t and $\xi$, $\textrm{F}^{Pom}_{prot}$ =  
$\textrm{F}^{Pom}_{prot} (t,\xi)$, with $t$ the 4-momentum transfer to the 
proton, and $\xi$ the fractional longitudinal momentum loss of the proton, 
$\xi$ = 1.$-$x$_{F}$.

\noindent An analysis of the Pomeron structure function arrives at the 
distribution of Pomerons in the proton parameterised in the following 
form (averaged over azimuthal angle $\phi$) \cite{DL}

\begin{equation}
\textrm{F}\textrm{$^{Pom}_{prot}(t,\xi)$} = 
\textrm{$\frac{9 \beta_{0}^{2}}{4\pi^{2}}$}\;
\textrm{$[F_{1}(t)\Large]^{2}$}\textrm{$\xi^{1-2\alpha(t)}$},
\label{eq5}
\end{equation}
 
\noindent with $\beta_{0}$ = 1.8 GeV$^{-1}$, $F_{1}$(t) the elastic form factor,
and $\alpha$(t) the Pomeron trajectory  
$\alpha(t)\!=\!1.\!+\!\varepsilon\!+\!\alpha^{'}t$ 
with $\varepsilon \sim 0.085$, \mbox{$\alpha^{'}\!$ = $\!$ 0.25 GeV$^{-2}$.}

\noindent The total cross section for exclusive resonance production at 
hadron level is expressed as

\begin{equation}
\sigma_{pp} = 
\!\!\!\!\int\!\!\!\!\int\!\!\!\!\int\!\!\!\!\int\!\!\!\!\int\!\!\!\!\int\! 
\frac{\textrm{flux}_{Pom}}{\textrm{flux}_{prot}} \cdot 
F^{Pom}_{prot_{A}}(t_{A},\xi_{A},\phi_{A}) 
\;F^{Pom}_{prot_{B}}(t_{B},\xi_{B},\phi_{B})\;\sigma_{PP}(M_{x},t_{A,B})\;
dt_{A}d\xi_{A}d\phi_{A}dt_{B}d\xi_{B}d\phi_{B}.
\label{eq6}
\end{equation}

\noindent With a kinematic transformation 
$(t_{A},\xi_{A},t_{B},\xi_{B},\Delta\phi)\!\longmapsto\! 
(u_{+},u_{-},v_{+}, M_{x}, p_{T,x})$ 
with Jacobian $J$, the cross section becomes

\begin{equation}
\sigma_{pp} =
\!\!\!\int\!\!\!\!\int\!\!\!\!\int\!\!\!\!\int\!\!\!\!\int\!
\frac{\textrm{flux}_{Pom}}{\textrm{flux}_{prot}} 
\cdot \tilde{F}^{Pom}_{prot_{A}} \tilde{F}^{Pom}_{prot_{B}}
\frac{p_{T,x}dp_{T,x}}{\sqrt{F^{2}\!-(p^{2}_{T,x}\!-\!G)^{2}}} 
\frac{\sigma_{PP}(M_{x},u_{+},u_{-})M_{x} \; J dM_{x}du_{+}du_{-}dv_{+}}
{\sqrt{H^{2}-(\frac{p^{2}_{T,x}+M^{2}_{x}}{2\gamma^{2}})}},
\label{eq7}
\end{equation}

\noindent with $M_{x}$ and $p_{T,x}$ the mass and transverse momentum of the 
meson state, respectively, and $F, G$ and $H$ known functions of the new 
variables $u_{+},u_{-},v_{+}$.

\noindent From Equation (\ref{eq7}), the double differential cross section 
is defined as 
\begin{equation}
\frac{d\sigma_{pp}}{dM_{x}dp_{T,x}} =
\!\!\!\int\!\!\!\!\int\!\!\!\!\int\!
\frac{\textrm{flux}_{Pom}}{\textrm{flux}_{prot}} \cdot \tilde{F}^{Pom}_{prot_{A}} 
\tilde{F}^{Pom}_{prot_{B}}\frac{p_{T,x}}{\sqrt{F^{2}\!-(p^{2}_{T,x}\!-\!G)^{2}}}
\frac{\sigma_{PP}(M_{x},u_{+},u_{-})M_{x} \; J du_{+}du_{-}dv_{+}}
{\sqrt{H^{2}-(\frac{p^{2}_{T,x}+M^{2}_{x}}{2\gamma^{2}})}}.
\label{eq8}
\end{equation}

\begin{figure*}[h]
\begin{center}
\vspace{-0.2cm}
\includegraphics[width=.38\textwidth]{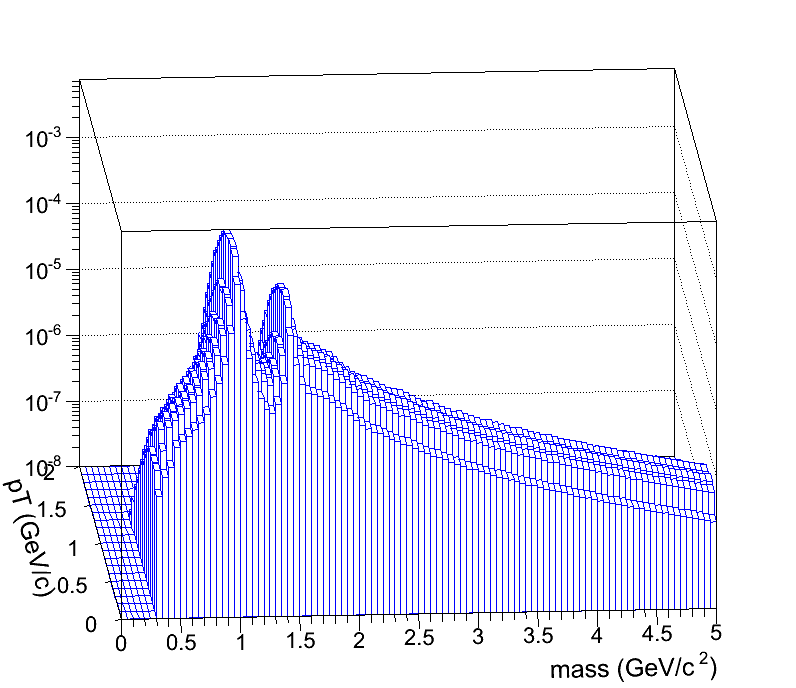}
\hspace{0.6cm}
\includegraphics[width=.38\textwidth]{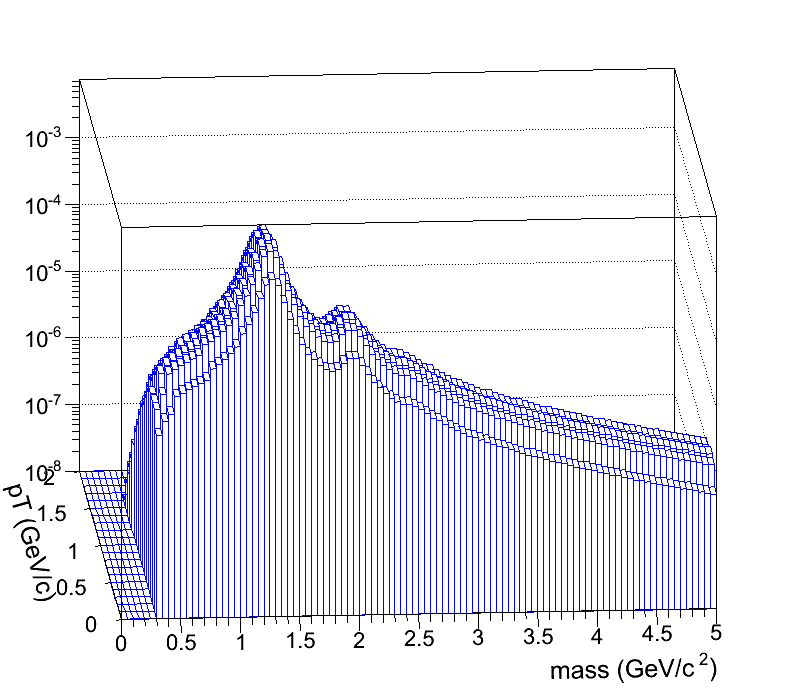}
\caption{The differential cross section for resonance production by 
$f_{1}$ trajectory on the left, and by $f_{2}$ trajectory on  the right.}
\label{fig3}
\end{center}
\end{figure*}

\noindent The triple differential cross section 
$\frac{d\sigma(pp \rightarrow pp f_{0}(980))}{dM_{x} dp_{T,x} dy}$ is shown in 
units of mb$\!$ GeV$^{-2}c^{3}$ on the left of Figure \ref{fig3} for resonance 
production by $f_{1}$ trajectory at midrapidity. The corresponding 
differential cross section for the $f_{2}$ trajectory is shown on the right.

\begin{figure*}[h]
\begin{center}
\includegraphics[width=.42\textwidth]{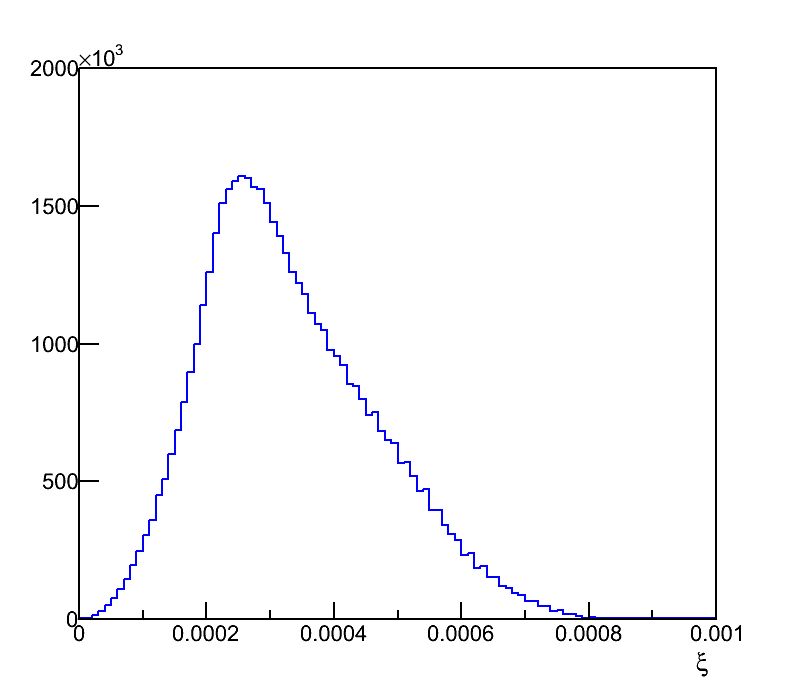}
\caption{Proton fractional longitudinal momentum loss for 
f$_{0}$(980) production.}
\label{fig4}
\end{center}
\end{figure*}

The proton fractional longitudinal momentum loss $\xi$ for f$_{0}$(980) 
production at a  center-of-mass \mbox{energy $\sqrt{s}$ = 14 TeV} is shown 
in Figure \ref{fig4}.  Values of $\xi$ in the range 1-6 $\times 10^{-4}$ are 
reached. Such low $\xi$-values underline the interest of central diffractive 
production studies in the low-mass region for examining the Pomeron 
distribution in the proton at low values of Bjorken-$x$. In this kinematical 
range, the onset of gluon saturation is one  of the central issues for 
studying QCD induced dynamics at the LHC.

\section{ACKNOWLEDGMENTS}
This work is supported by the German Federal Ministry of Education and 
Research under promotional reference 05P15VHCA1.


\nocite{*}
\bibliographystyle{aipnum-cp}%
\bibliography{sample}%

\end{document}